\documentclass[twocolumn,showpacs]{revtex4}
\usepackage{graphicx}
\usepackage{epstopdf}
\usepackage{bm}

\begin{document}
\title{Resonantly enhanced tunneling of Bose-Einstein condensates in periodic potentials}

\author{C. Sias, A. Zenesini, H. Lignier, S. Wimberger, D. Ciampini, O. Morsch and E. Arimondo}
\affiliation{CNR-INFM, Dipartimento di Fisica `E. Fermi', Largo
Pontecorvo 3, 56127 Pisa, Italy}

\begin{abstract}
We report on measurements of resonantly enhanced tunneling of
Bose-Einstein condensates loaded into an optical lattice. By
controlling the initial conditions of our system we were able to
observe resonant tunneling in the ground and the first two excited
states of the lattice wells. We also investigated the effect of
the intrinsic nonlinearity of the condensate on the tunneling
resonances.

\end{abstract}


\maketitle

Resonantly enhanced tunneling (RET) is a quantum effect in which
the probability for tunneling of a particle between two potential
wells is increased when the quantized energies of the initial and
final states of the process coincide. In spite of the fundamental
nature of this effect~\cite{bohm} and the practical
interest~\cite{chang}, it has been difficult to observe
experimentally in solid state structures. Since the 1970s, much
progress has been made in constructing solid state systems such as
superlattices~\cite{chang_app,esaki_review,glutsch04} and quantum
wells~\cite{wagner93} which enable the controlled observation of
RET~\cite{leo03}.

In recent years, ultra-cold atoms in optical
lattices~\cite{grynberg} have been increasingly used to simulate
solid state systems. Optical lattices are easy to realize in the
laboratory, and their parameters can be perfectly controlled both
statically and dynamically. Also, more complicated potentials can
be realized by adding further lattice beams~\cite{santos}. This
makes them attractive as model systems for crystal lattices, and
in the last few years cold atoms and Bose-Einstein condensates
(BECs) in optical lattices have been used to simulate phenomena
such as Bloch oscillations~\cite{morsch_review} and the Mott
insulator transition~\cite{greiner}. In this Letter we show that
BECs in accelerated optical lattice potentials are ideally suited
to studying RET. While in solid state measurements of RET only a
few potential wells were used and the periodic structures had to
be grown for each realization, in our experiment the condensate is
distributed over several tens of wells and the parameters of the
lattice can be freely chosen. Moreover, we are able to control the
initial conditions of the system and thus observe RET in any
chosen energy level and can also add nonlinearity to the system.

\begin{figure}[ht]
\includegraphics[width=8cm]{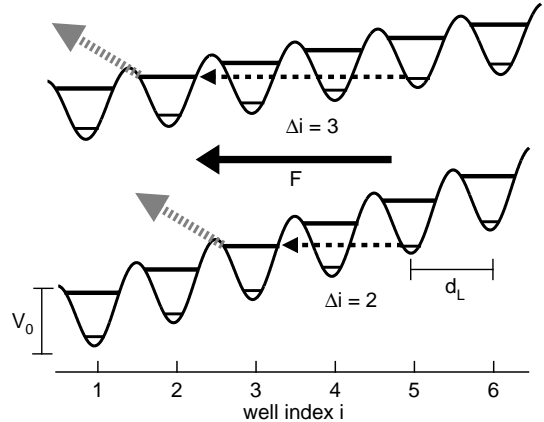}
\caption{\label{figure1} Explanation of resonantly enhanced
tunneling. Tunneling of atoms out of a tilted lattice is
resonantly enhanced when the energy difference between lattice
wells matches the distance between the energy levels in the
wells.}
\end{figure}

\begin{figure}[ht]
\includegraphics[width=8cm]{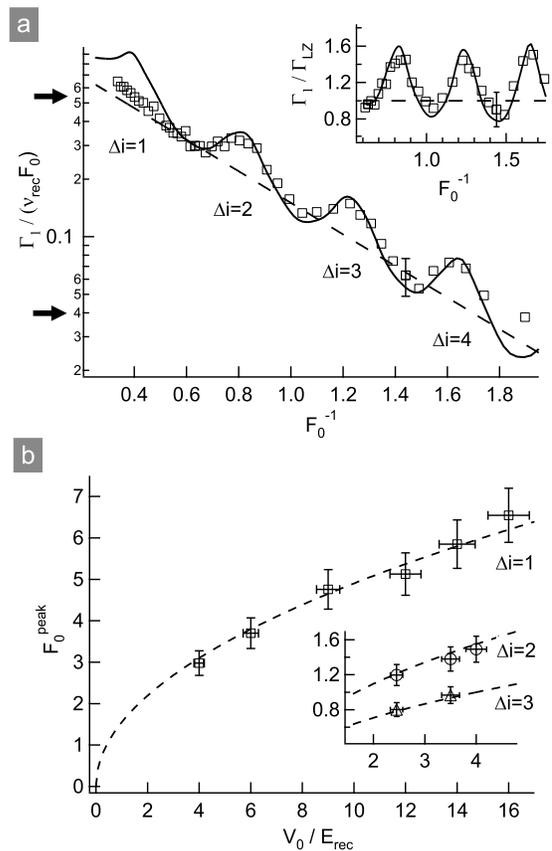}
\caption{\label{figure2} Tunneling resonances in an accelerated
optical lattice. (a) Tunneling resonances of the $n=1$ lowest
energy level for $V_0=2.5\,E_{rec}$. The arrows indicate the upper
and lower limits for our precise measurement of $\Gamma_n$. Inset:
Deviation from the Landau-Zener prediction. For clarity, in both
graphs only one representative error bar is shown. (b) Positions
of the $\Delta i=1$ resonance peaks as a function of the lattice
depth. Only data points for which the resonance is clearly visible
(e.g. not $\Delta i =1$ of (a)) are included. Inset: Positions of
the peaks for $\Delta i =2$ and $3$.}
\end{figure}

A schematic representation of RET is shown in Fig. 1. In a tilted
periodic potential, atoms can escape by tunneling to the continuum
via higher-lying levels. The tilt of the potential is proportional
to the force $F$ acting on the atoms, and in general the tunneling
rate $\Gamma_{LZ}$ can be calculated using the Landau-Zener
formula~\cite{zener32}. However, when the tilt-induced energy
difference $Fd_L\Delta i$ between wells $i$ and $i+\Delta i$
matches the separation between two quantized energy levels, the
tunneling probability is resonantly enhanced and the Landau-Zener
formula no longer gives the correct result, as previously
investigated in~\cite{bharucha} for cold atoms in optical
lattices. While for the parameters of our experiment the
enhancement over the Landau-Zener prediction was around a factor
of $2$ (see theoretical and experimental results of Fig.
\ref{figure2}(a)), in general it can be several orders of
magnitude.

The starting point of our experiments is a BEC of
$^{87}\mathrm{Rb}$ atoms, held in an optical dipole trap whose
frequencies can be adjusted to realize a cigar-shaped condensate.
The BECs are created using a hybrid approach in which evaporative
cooling is initially effected in a magnetic time-orbiting
potential (TOP) trap and subsequently in a crossed dipole trap.
The dipole trap is realized using two intersecting gaussian laser
beams at $1030\,\mathrm{nm}$ wavelength and a power of around
$1\,\mathrm{W}$ per beam focused to waists of $50\,\mathrm{\mu
m}$. After obtaining pure condensates of around $5\times 10^4$
atoms the powers of the trap beams are adjusted in order to obtain
an elongated condensate with the desired trap frequencies
($\approx 20\,\mathrm{Hz}$ in the longitudinal direction and
$80-250\,\mathrm{Hz}$ radially).

Subsequently, the BECs held in the dipole trap are loaded into an
optical lattice created by two gaussian laser beams ($\lambda =
852\,\mathrm{nm}$) with $120\,\mathrm{\mu m}$ waist intersecting
at an angle $\theta$. The resulting periodic potential
$V(x)=V_0\sin^2(\pi x/d_L)$ has a lattice spacing $d_L= \lambda /
( 2\sin(\theta /2))$ and its depth $V_0$ is measured in units of
the recoil energy $E_{\rm rec}= \hbar^2 \pi^2 / (2m d_L^2)$, where
$m$ is the mass of the Rb atoms.  In the present experiment, we
used $d_L=0.426\,\mathrm{\mu m}$ (for $V_0/E_{\rm
rec}=6,4,9\,\mathrm{and}\, 16$) and $d_L=0.620\,\mathrm{\mu m}$
(for $V_0/E_{\rm rec}=2.5,10,12\,\mathrm{and}\, 14$). By
introducing a frequency difference $\Delta \nu$ between the two
lattice beams (using acousto-optic modulators which also control
the power of the beams), the optical lattice can be moved at a
velocity $v=d_L\Delta \nu$ or accelerated with an acceleration
$a=d_L\frac{d\Delta\nu}{dt}$.

A ramp from $0$ to $V_0$ in around $1\,\mathrm{ms}$ loads the BEC
adiabatically into the optical lattice~\cite{footnote2}. For
loading the ground state levels, the lattice velocity is $v=0$
during the ramp. For the first and second excited levels, during
the ramp the lattice is moved at a finite velocity calculated from
the conservation of energy and quasimomentum~\cite{peik97}.
Finally, the optical lattice is accelerated with acceleration $a$
for an integer number of Bloch oscillation cycles. In the rest
frame of the lattice, this results in a force $F=ma$ on the
condensate. Atoms that are dragged along by the accelerated
lattice acquire a larger final velocity than those that have
undergone tunneling, and are spatially separated from the latter
by releasing the BEC from the dipole trap and lattice at the end
of the acceleration period and allowing it to fall under gravity
for $5-20\,\mathrm{ms}$. After the time-of-flight, the atoms are
detected by absorptive imaging on a CCD camera using a resonant
flash.

From the dragged fraction $N_{drag}/N_{tot}$, we then determine
the tunneling rate $\Gamma_n$ in the asymptotic decay law
\begin{equation}
N_{drag}(t)=N_{tot}\exp{(-\Gamma_n t)} \end{equation} where the
subscript $n$ indicates the dependence of the tunnnelling rate on
the local energy level $n$ in which the atoms are initially
prepared (ground state: $n=1$, first excited state: $n=2$, etc.).
In the experiments reported in this work, the number of bound
states in the wells was small (2-4, depending on the lattice
depth), so after the first tunneling event, the probability for
tunneling to the next bound state or the continuum was close to
unity.

The resolution of our tunneling measurement is given by the
minimum number of atoms that we can distinguish from the
background noise in our CCD images, which varies between $500$ and
$1000$ atoms, depending on the width of the observed region. With
our condensate number, and taking into account the minimum
acceleration time limited by the need to spatially separate the
two fractions after time-of-flight and the maximum acceleration
time limited by the field of view of the CCD camera, this results
in a maximum $\Gamma_n/\nu_{\mathrm rec}$ of $\approx 1$ and a
minimum of $\approx 1\times 10^{-2}$, with the recoil frequency
$\nu_{\mathrm{rec}}=E_{\mathrm{rec}}/h$.


A typical plot of the tunneling rate $\Gamma_{1}$ out of the
ground state as a function of $F_0^{-1}$ (where $F_0= F d_L/E_{\rm
rec}$ is the dimensionless force)
 in the
linear regime is shown in Fig. 2(a). This regime is reached either
by choosing small radial dipole trap frequencies or by releasing
the BEC from the trap before the acceleration phase and thus
letting it expand. In both cases, the density and hence the
interaction energy of the BEC is reduced. Superimposed on the
overall exponential decay of $\Gamma_1/F_0$ with $F_0^{-1}$, one
clearly sees the resonant tunneling peaks corresponding to $\Delta
i=2, 3$ and $4$ (for this choice of parameters, the $\Delta i=1$
peak lay outside our experimental resolution). In order to
highlight the deviation from the Landau-Zener prediction, in the
inset of Fig. \ref{figure2} (a) we plot $\Gamma_1/\Gamma_{LZ}$,
where the Landau-Zener tunnelling rate $\Gamma_{LZ}$ is given
by~\cite{zener32,peik97}
\begin{equation}
\Gamma_{LZ}=\nu_{\mathrm rec} F_0
e^{-\frac{\pi^2(V_0/E_{\mathrm{rec}})^2}{32F_0}}\,.
\end{equation} The experimental results
are in good agreement with numerical solutions obtained by
diagonalizing the Hamiltonian of the open decaying
system~\cite{gluck99,gluck02}. Figure~\ref{figure2}(b) summarizes
our results for the positions of the ground-state resonances
$\Delta i=1, 2$ and $3$ as a function of the lattice depth
together with a theoretical fit assuming the separation of the
lowest energy levels to be
\begin{equation}\Delta E = \alpha E_{\mathrm{rec}}
\sqrt{V_0/E_{\mathrm{rec}}}\,.\end{equation} Independently of
$\Delta i$, the best fit is achieved for $\alpha =1.5$, to be
compared with $\alpha=2$ for the harmonic oscillator
approximation. A value $\alpha<2$ is to be expected since our
lattice wells only contain a few bound states and are, therefore,
highly anharmonic.

\begin{figure}[ht]
\includegraphics[width=8cm]{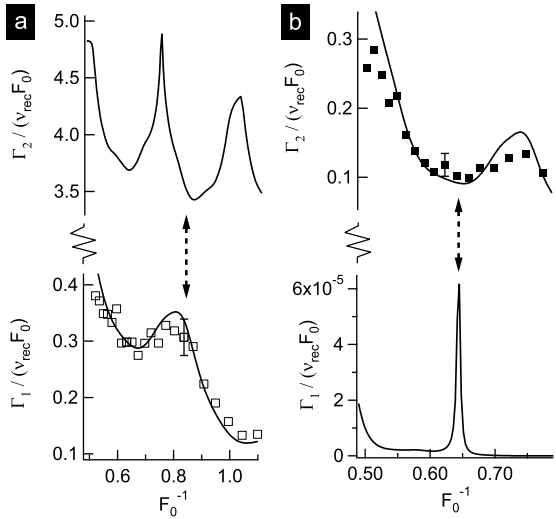}
\caption{\label{figure3} Anti-crossing scenario of the RET rates.
(a) Theoretical plot of $\Gamma_{1,2}$ for
$V_0=2.5\,E_{\mathrm{rec}}$ with experimental points for
$\Gamma_1$. (b) Theoretical plot of $\Gamma_{1,2}$ for
$V_0=10\,E_{\mathrm{rec}}$ with experimental points for
$\Gamma_2$. For clarity, the vertical axes have been split and the
$\Gamma_n$ plotted on a linear scale, and only one representative
error bar is shown.}
\end{figure}

Using BECs in optical lattices allows us to explore resonant
tunneling in regimes that are difficult or even impossible to
access in solid state systems. First, we can prepare the
condensates in the  excited levels of the lattice wells before the
acceleration. Again, tunneling resonances are clearly visible, and
the experimental results agree with theoretical calculations. The
accessibility of higher energy levels allows us to experimentally
determine the decay rates at resonance of two strongly coupled
levels. Although our experimental resolution does not allow us to
measure the decay rates in two different levels for the same set
of parameters $F_0$ and $V_0$, we are able to compare the ground
and excited state decay rates $\Gamma_1$ and $\Gamma_2$ with the
theoretical predictions for two different parameter sets, as shown
in Fig. 3. This figure reveals the anti-crossing of the decay
rates of strongly coupled levels as a function of our control
parameter $F_0$. These results demonstrate a peculiar behaviour of
the Wannier-Stark states studied
theoretically~\cite{avron82,wagner93} and more recently rephrased
within a general context of crossings and anti-crossings for the
real and imaginary parts of the eigenvalues of a non-hermitian
Hamiltonians~\cite{keck03}. Our data confirm the predictions of
\cite{gluck99} that the anti-crossings modify the decay rates of
the two perturbing states in different ways.

\begin{figure}[ht]
\includegraphics[width=8cm]{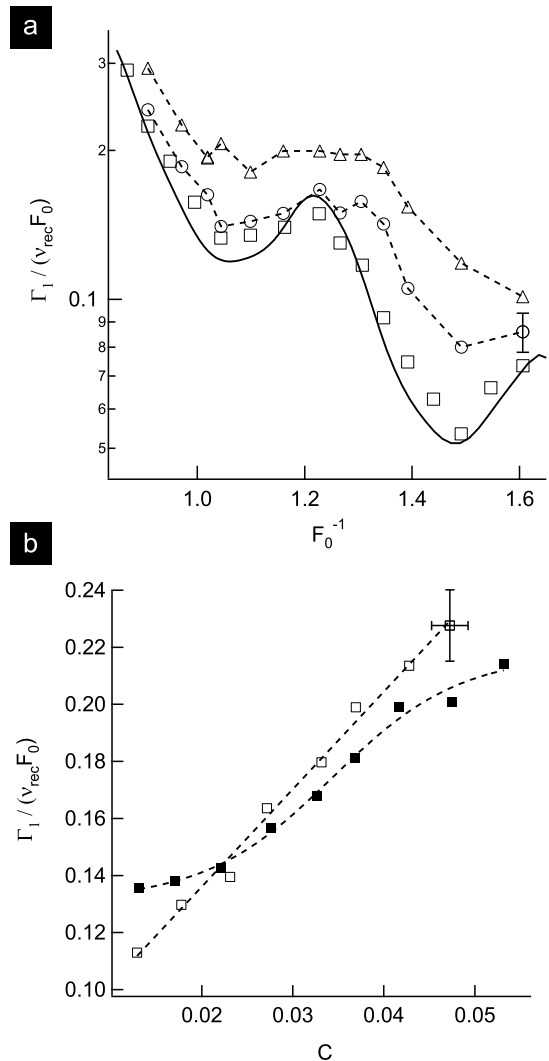}
\caption{\label{figure4} Resonant tunneling in the nonlinear
regime. (a) Resonance $\Delta i=3$ for $V_0=2.5\,E_{rec}$ with
$C=0.024$ (squares), $C=0.035$ (circles) and $C=0.057$
(triangles). The solid line is the theoretical prediction for
$C=0$; the dashed lines are guides to the eye. (b) Dependence on
$C$ of the tunneling rate at the position of the peak $F_0^{-1} =
1.21$ (solid symbols) and of the trough $F_0^{-1} = 1.03$ (open
symbols). The dashed lines are fits to guide the eye. For clarity,
in (a) and (b) only one typical error bar is shown.}
\end{figure}

Additionally, by exploiting the intrinsic nonlinearity of the
condensate due to atom-atom interactions, we can study RET
 in the nonlinear regime, as simulated in~\cite{wimberger05}.
 In order to realize this
regime, we carry out the acceleration experiments in radially
tighter traps (radial frequency $\gtrsim 100\,\mathrm{Hz}$) and
hence at larger condensate densities. Figure 4(a) shows the
results for increasing values of the nonlinear
parameter~\cite{morsch01}
\begin{equation}C= \frac{n_0 a_sd_L^2}{\pi},
\end{equation} where $n_0$ is
the peak condensate density and $a_s$ the $s$-wave scattering
length. Two effects are visible:
 First,  the overall (off-resonant)  level of $\Gamma_1$
increases linearly with $C$. This is in agreement with our earlier
experiments on nonlinear Landau-Zener
tunneling~\cite{morsch01,jonalasinio03} and can be explained describing the condensate evolution within  a
nonlinearity-dependent effective potential
$V_{\mathrm{eff}}=V_0/(1+4C)$~\cite{choiniu99}. Second, with
increasing nonlinearity, the contrast of the RET peak is decreased
and the peak eventually vanishes. This is confirmed by the
different dependence on $C$ of the on- and off-resonant values of
$\Gamma_1$ (Fig. 4 (b)). We estimate that in order to
significantly affect the resonant tunneling rate, the nonlinearity
parameter has to be comparable to the width of the RET peak. This
order-of-magnitude argument agrees with our observations.

Finally, we have experimentally tested the robustness of RET
against a dephasing of the lattice wells induced by non-adiabatic
loading of the BEC into the lattice in the nonlinear
regime~\cite{morsch_decay,gericke06}. Even for completely dephased
wells, the tunneling resonances survive.

In summary, we have measured resonantly enhanced tunneling of BECs
in accelerated periodic potentials in a regime where the standard
Landau-Zener description is not valid. Our results in the linear
regime agree with numerical calculations, and the possibility to
observe RET for arbitrary initial conditions and parameters of the
periodic potential underlines the advantage of our system over
solid state realizations. Furthermore, we have explored RET in the
nonlinear regime and demonstrated that, as theoretically
predicted, the tunneling resonances disappear for large values of
the nonlinearity.

In the present set-up the measurement of the tunneling rate is
limited in its dynamic range by the detection geometry. A larger
dynamic range can be realized by long-distance transport of
BECs~\cite{schmid06}. Our method for observing RET can also be
generalized in order to study other regular or disordered
potentials, the effects of noise and the presence of a thermal
fraction in the condensate. Furthermore, one might exploit the
tunneling resonances to explore the spatial decoherence processes
and to perform precision measurements.

This work was supported by the European Community STREP Project
OLAQUI, a MIUR-PRIN Project, the Sezione di Pisa dell'INFN, and
the Feodor-Lynen Programme of the Alexander v. Humboldt
Foundation. The authors would like to thank M. Cristiani, R.
Mannella and Y. Singh for assistance, A. Kolovsky for useful
discussions and S. Rolston for a critical reading of the
manuscript.

\bibliographystyle{apsrmp}

\begin{thebibliography}{99}
\bibitem{footnote2} In the nonlinear regime, i.e. when the radial
trap frequency is large and hence the interaction energy of the
BEC is appreciable, the speed of the ramp has to be reduced and
loading times can be as long as tens of
milliseconds~\cite{gericke06}.
\end{thebibliography}

\end{document}